\documentclass[12pt]{article}
\usepackage{epsf}
\title{Moments of the heavy-quark parton distribution
function from QCD sum rules}
\vspace{2cm}
\author{A.G.Oganesian\\
\\
Institute of Theoretical and Experimental Physics\\
B.Cheremushkinskaya 25, 117218, Moscow, Russia}
\date{}
\begin{document}
\maketitle

\date{}

\maketitle

\newcommand{\be}{\begin{equation}}
\newcommand{\ee}{\end{equation}}

\def\la{\mathrel{\mathpalette\fun <}}
\def\ga{\mathrel{\mathpalette\fun >}}

\def\fun#1#2{\lower3.6pt\vbox{\baselineskip0pt\lineskip.9pt

\ialign{$\mathsurround=0pt#1\hfil##\hfil$\crcr#2\crcr\sim\crcr}}}

\begin{abstract}

The moments of the heavy quark-parton distribution functions
in a heavy pseudoscalar meson are calculated from QCD sum rules.
Expanding these sum rules in the inverse heavy quark mass we obtain
the heavy-mass limits of the moments. Comparison with the finite
mass results reveals that while the heavy mass expansion works
reasonably well for the $b$ quark, one has to take into account
terms of higher than $(1/m_c)^2$ order for the $c$ quark.
This result can provide a quantitative assessment of
$c$ and $b$ quark fragmentation models based on
the heavy-quark mass limit.
\end{abstract}

\section{}

The knowledge of the heavy quark fragmentation functions is very
important for description of various processes of
heavy hadron production. There exist many  different theoretical models and
approaches, in  which these functions  are parameterized
and their perturbative part calculated in QCD (see, for example
\cite{a1}-\cite {a4}). Some of  these models (e.g. \cite{a4})
use  expansions, based on the heavy-quark mass limit or on HQET.
That is why it seems interesting to reconsider the
moments of the heavy-quark fragmentation functions,
obtained long ago from QCD sum rules in \cite{n1}.
In this approach the moments of the heavy-quark fragmentation functions
into heavy-light meson (e.g., the $c$-quark fragmentation into a $D$ meson)
\be
M^D_n=\int\limits_0^1 dz z^{n-1} D^c(z,Q^2)
\ee
and the moments of heavy-quark
parton distribution functions in the same meson
\be
M_n^c=\int\limits_0^1 dx x^{n-1} c(x,Q^2)
\ee
were equated (in the scaling limit) following the familiar
relation \cite {gl} ( see also \cite{kart} and \cite{n1} for details).
The moments $M_n^c$ were directly calculated from QCD sum rules.
The advantage of this method is the possibility  to estimate
the heavy parton distribution functions (hence, also fragmentation
functions) in full QCD, in terms of universal parameters, such as quark masses
and condensate densities. One can use this result to fix the parameters of
different models, as was discussed in \cite{n1}.

In addition, as will be shown here it is possible
to check whether the expansion in the inverse heavy-quark mass is reliable
for the description of the fragmentation functions, especially in
the case of the c-quark. In what follows I address this question
considering the fragmentation of heavy quarks to pseudoscalar
heavy-light meson. A more complete study including
mesons with different quantum numbers and assessing the
heavy-quark spin symmetry violation in the fragmentation functions
will be presented in the paper being in preparation.

Let me briefly remind the method and results of the paper
\cite{n1}. For definiteness the $c$-quark fragmentation to
$D$ meson is considered, hence one needs the QCD sum rule
for $c$ quark-parton distribution in $D$-meson.
One starts from the four-point correlator:
%\be
%\Pi_{\mu\nu\lambda\rho}(p_1,p_2;q_1,q_2) = i\int
%e^{ip_1x+iq_1y-ip_2z}d^4xd^4yd^4z\langle 0\mid T \left
%\{j_{\lambda}(x) j^{em}_{\mu}(y)j^{em}_{\nu}(0)j_{\rho}(z)\right
%\}\mid 0 \rangle
%\ee
\be
\Pi_{\mu\nu}(p_1,p_2;q_1,q_2) = i\int
e^{ip_1x+iq_1y-ip_2z}d^4xd^4yd^4z\langle 0\mid T \left \{j_5(x)
j^{em}_{\mu}(y)j^{em}_{\nu}(0)j_5^\dagger(z)\right \}\mid 0 \rangle
\ee
where the two $c$-quark electromagnetic
currents, $j^{em}_{\mu}=\bar{c}\Gamma_{\mu}c $,
are correlated with the two heavy-light pseudoscalar
currents $j_5=\bar{c}\Gamma_5u $ that interpolate $D$ meson:
%$ j_{\lambda}=\bar{c}\Gamma_{\lambda}u $,
%$j_{\rho}=\bar{u}\Gamma_{\rho}c $

The above correlator is considered in the deep spacelike region at
$t=(p_1-p_2)^2=0$, $p_i^2,q_i^2<0$, where all internal quarks
are highly virtual and the operator-product expansion (OPE)
is applicable. We take into account the contributions of the unit
operator (bare loop), quark condensate
$\langle 0\mid \bar{\psi}\psi\mid 0\rangle$ and
quark-gluon condensate $\langle 0\mid
\bar{\psi}G_{\mu\nu}\sigma^{\mu\nu}\psi\mid 0\rangle$. The
two latter contributions are important, their
coefficients being proportional to the heavy quark mass.
The gluon- and four-quark-condensate contributions are neglected,
because the estimates show, that they are very small.

Importantly, the correlator is calculated
as a function of two independent external momenta squared $p_1^2,p_2^2$.
In this way one can reliably extract the ground state
$D$-meson contribution both in the initial and final states, making
the resulting sum rules more accurate. Hereafter we will consider the
invariant amplitude multiplying the kinematical structure $g_{\mu\nu}$:
$\Pi_{\mu\nu}= g_{\mu\nu}\Pi + ...$. Let us express this amplitude
in terms of double dispersion relation over $p_1^2$ and $p_1^2$, i,e:
\be \Pi=
\frac{1}{\pi^2}\int\frac{du_1}{u_1-p_1^2}\int\frac{du_2}{u_2-p_2^2}\,
\rho(u_1,u_2,s)
\label{1}
\ee
We will consider the derivatives of the amplitude $\Pi$ over
$s=(p_1+q_1)^2$. Furthermore, using the dispersion representation over s,
one can write the spectral density $\rho$ in the form
\be
\frac{1}{n!}\frac{d^n}{ds^n}\rho(u_1,u_2,s)=
\int\frac{ds_1}{(s_1-s)^{n+1}}f^0(u_1,u_2,s_1)\,.
\label{2}
\ee
As usual in the derivation of a QCD sum rule,
one should saturate the amplitude (\ref{2})
by hadronic states and equate it to the result of OPE
(the bare loop and operators  of dimensions 3 and 5, as
mentioned above).
One easily finds  the contribution of the ground ($D$-meson) states in
the initial and final hadron channels  to the equation (\ref{2}):
\be \rho=Im_{u_1}Im_{u_2}\Pi=-\pi^2 g_D^2 m_D^4
\delta(u_1-m_D^2)\delta(u_2-m_D^2)T_1
\label{4}
\ee
where $g_D m_D^2= i\langle 0\mid \bar{c}\gamma_5c\mid c\rangle$
and $T_1$ is defined by following relation
\be \int e^{iq_1x}d^4x\langle D\mid T
\left\{j^{em}_{\mu}(x)j^{em}_{\nu}(0)\right \}\mid D \rangle  =
g_{\mu\nu}T_1 + ...
\label{5}
\ee
The above amplitude corresponds to the amplitude of the
virtual photon scattering on $D$ meson and is therefore
obviously related with the c-quark parton distribution
(structure function of $D$ meson) by well known relation
\be Im_sT_1=-\pi c(x).
\label{6}
\ee
Now one should substitute eq. (\ref{4}-\ref{6}) into
eq. \ref{2}, then perform  two
independent Borel transformation on $p_1^2$  and $p_2^2$ , with the
Borel  masses $M_{1Bor}^2$ and $M_{2Bor}^2$, respectively. We
will equate the Borel masses $M_{1Bor}^2 = M_{2Bor}^2 =M^2$  only in
the final sum rule. Finally, the hadronic part of the sum rule
can now be written as
\begin{eqnarray}
\frac{1}{n!}\frac{d^n}{ds^n}\Pi= -\pi^2 g_D^2
m_D^4exp[-2m_D^2/M_{Bor}^2]\frac{1}{\pi}
\int\limits_{m_c^2}\frac{ds_1}{(s_1-s)^{n+1}}Im_sT_1(Q^2,s1)
\nonumber\\
=-\pi^2 g_D^2 m_D^4exp[-m_D^2/M_{Bor}^2]\frac{1}{\pi}
\int\limits^1_0\frac{dx}{x^2}C(x,Q^2)(m_D^2-Q^2+Q^2/x-s)^{-n-1}=
\label{7}
\end{eqnarray}
Here we adopt quark-hadron duality approximation for
the hadron continuum contribution, so that it cancels
with the appropriate part of the bare loop contribution.
Following \cite{n1}, we choose the value of the variable
$s$ in the unphysical region as $s=m_D^2-Q^2$, and
transform the integrals in the first line of eq.(\ref{7})
into  the moments of the parton distribution.
of the $c$ quark in $D$ meson.
Finally, equating the hadronic  representation (7)
to the OPE result, one can write down the following sum
rules for these moments:
\begin{eqnarray}
M_n=g_D^{-2}m_D^{-4}\Biggl ( \frac{1}{\pi^2}\int\limits^{s_0}_{m^2}du_1exp\Biggl[\frac{-2(u_1-m_D^2)}{M^2}\Biggr]
\int\limits^{z_+}_{z_-}dz f_0 *(1+z-m_D^2/Q^2)^{-n-1} \nonumber\\
+\frac{ma}{4\pi^2}L\eta^{-n-1}exp\Biggl[\frac{-2(m^2-m_D^2)}{M^2}\Biggr]
+ R_6\Biggr)
\end{eqnarray}
\label{8}
where
\be
f_0=\frac{3}{8}(s_1-m^2)\Biggl[
\Biggl( 1+z+\frac{s_1+2m^2}{Q^2} \Biggr)
\Biggl((1+z-s_1/Q^2)^2+\frac{4s_1}{Q^2}\Biggr )^{-1/2}
-\frac{4m^2}{Q^2}\Biggl (1+\frac{4m^2}{Q^2}\Biggr )^{-1/2}\Biggr
]
\label{9}
\ee
and

\begin{eqnarray}
R_6 = \frac{m m_0^2 a}{4\pi^2}\frac{1}{2M^2}\Biggl
[4/M^2-2m^2/M^4 +1/Q^2+ \frac{(n+1)}{\eta}\Biggl(\frac{2}{3Q^2}
-1/M^2 - \frac{2m^2}{M^2Q^2}\Biggr)
\nonumber\\
-\frac{(n+1)(n+2)m^2}{Q^4\eta^2} \Biggr ]
\eta^{-n-1}exp\Biggl[\frac{-2(m^2-m_D^2)}{M^2}\Biggr]
\label{10}
\end{eqnarray}
where
$$z_+=u_1(1/Q^2+1/(2m^2))-1/2+ (1+4m^2/Q^2)^{1/2}(u_1/m^2-1)/2$$\,,

$$z_-=u_1(1/Q^2+1/(2m^2))-1/2- (1+4m^2/Q^2)^{1/2}(u_1/m^2-1)/2$$\,.

Here $a=- (2\pi)^4\langle 0\mid\bar{\psi}\psi\mid 0\rangle$,
$\eta=(1-(m^2-m_D^2)/Q^2)$. The factor
$L=ln(m/\lambda)/ln(\mu\lambda))^{4/9}$ accounts for the anomalous
dimension of the quark condensate, normalized at point $\mu$.
We adopt the values of condensates \cite{giz}, \cite{iof},
normalizing them  at $\mu=1 \mbox{GeV}$. In this case the
factor $L$ is close to the unity.

In \cite{n1},  it was shown that the first few moments of the sum
rules are well behaved and the following numerical results were
obtained:

$$M_2=0.85, M_3=0.75, M_4=0.67$$ at $Q_0^2=20GeV^2 $

$$M_2=0.9, M_3=0.83, M_4=0.78$$ at $Q_0^2=10GeV^2 $.

These moments can be use as an input in the evolution equations,
at some initial point $Q_0^2 = (3-5)m_D^2$, hence it is possible
to predict the structure functions also at large $Q^2$.

In what follows  we will use the sum rules on order to
address only one question, if it is
possible to use the heavy quark  limit for
the heavy quark parton distribution (and, correspondingly,
fragmentation function). For that,
we expand the sum rules  for the moments in the
inverse mass of  the heavy quark.

Substituting in eq(9) instead of the $g_D^2 m_D^4$  the sum rule
obtained from the two-point correlator  \cite{aliev}
\begin{eqnarray}
g_D^2m_D^4=exp(m_D^2/M^2_B)
\Biggl( \frac{3}{8\pi^2}\int\limits_{m^2}du (u-m^2)exp(-u/M^2_B)/s
\nonumber\\
+\frac{ma}{4\pi^2}\Biggl[L+\frac{m_0^2}{2M^2_B}(2-m^2/M^2)exp(-m^2/M^2_B)\Biggr]\Biggr)
\label{11}
\end{eqnarray}
One can then rewrite the sum rules for  $M_n$ as
\be
M_n=
\frac{R_0(n)+R_4(n)+R_6(n)}{K_0+K_4+K_6}
\label{12}
\ee
where the terms $R_d(n)$ in the numerator
and $K_d$  in the denominator originate from
the dimension $d=0,4,6$ contributions to the OPE of the four-point
and two-point correlator, respectively. Following \cite{n1}
we adopt the relation
$M^2_B=M^2/2$ between the Borel masses in the four-point
and two-point sum rules.
Note that the exponential factor containing $m_D^2$
cancels in the ratio of the two sum rules.

To investigate the heavy-quark mass limit $m\to \infty$ limit
of (\ref{12}) we employ the standard scaling relations
for the heavy hadron mass,
continuum threshold and Borel parameter:
\be
m_D=m+ \bar{\Lambda},~~
s_0=m^2+2m\omega,~~
M^2_B=2\tau m
\label{13}
\ee
and in addition assuming the scaling $Q^2=\gamma m_D^2$,
where in the heavy quark limit  the parameters
$\bar{\Lambda}$,$\omega$, $\tau$ and $\gamma$
do not depend of the heavy quark mass.

The terms entering the numerator of eq.(\ref{13})
have the following expressions:
\be
R_0 (n)=  \int\limits^{z_m}_0 \frac{z dz}{2\tau}e^{-z}I_0(n)
\label{3a} \ee
where
\be I_0(n) =  \int\limits^{t_+}_{t_-} \Biggl
[\frac{1+4(\rho/\gamma)+t/m
+4(\varphi/m)(\rho/\gamma)}{\sqrt{(1+t/m)^2+4(\rho/\gamma)
+8(\varphi/m)(\rho/\gamma)}} -\frac{4(\rho/\gamma)}{v} \Biggr
]\frac{dt}{\lambda^{n+1}}  \ee\,,
$\varphi=z\tau$, $\delta= \bar{\Lambda}/m$, $z_m=\omega/\tau$,
and
$$\rho= m^2/m_D^2=1/(1+2\delta+\delta^2)$$

$$\lambda= 1+t/m +(\rho -1)/\gamma +2(\varphi\rho)/m$$

$$v=\sqrt{1+4\rho/\gamma}$$

$$t_+=z\tau(1+v)$$

$$t_-=z\tau(1-v)$$

$$y=\bar{\Lambda}/\tau$$.

%For simplicity we do not write the common factor,
%$$\frac{3m\tau^3}{\pi^2}e^{y(1+\delta/\gamma)}$$
%which appear both in the nominator and
%denominator, because it will be cancelled in the ratio (12).
Expanding in the powers of $1/m$ , we
obtain the first three terms for $R_0$:
\begin{eqnarray}
R_0(n) \simeq D_2 +(n+1)\delta\frac{2D_2-(2+\gamma)D_3}{\gamma}
+\delta^2(n+1)(n+2) \Biggl  [
\frac{2(n+2)+3\gamma}{\gamma^2(n+2)}D_2
\nonumber\\
-
\frac{2}{\gamma}(1+\frac{2(n+1)}{(n+2)\gamma})D_3 +
\frac{2D_4}{3\gamma^2}(\gamma^2+4\gamma+3
+\frac{2\gamma}{(n+1)(n+2)}) \Biggr ]
\end{eqnarray}

Here $D_n$ is defined as
$$D_n=n!E_n(\omega/\tau)/y^{n-2}$$
and
$$E_n(z)=1-e^{-z}-ze^{-z}-z^2e^{-z}/(2!)-
-z^ne^{-z}/(n!)$$

For the contribution of $d=4 $ operators we obtain:
\be
R_4(n)=\frac{a}{12\tau^3}L\frac{1}{\kappa^{n+1}}\,,
\ee
where $\kappa=1+(\rho-1)/\gamma$.
After expansion the first three terms are:
\be R_4(n) \simeq \frac{a}{12\tau^3}L\Biggl ( 1+2\delta(n+1)/\gamma
+\delta^2\frac{(n+1)(n+2)}{\gamma}[2/\gamma-3/(n+2)]\Biggr ) \ee

Finally, the contribution of $d=6 $ operators transforms into
\begin{eqnarray}
R_6(n)=\frac{a}{12\tau^3}\frac{m_0^2}{16\tau^2} \Biggl [
-1+8\epsilon- 2\epsilon\frac{(n+1)}{\kappa}(1+2\rho/\gamma)
\nonumber\\
+8\epsilon^2 \Biggl (
\frac{\rho}{\gamma}+\frac{2(n+1)}{3\kappa}\frac{\rho}{\gamma}-\frac{(n+1)(n+2)}{2\kappa^2}\frac{\rho^2}{\gamma^2}
\Biggr )\Biggr ]\frac{1}{\kappa^{n+1}}\,,
\end{eqnarray}
where $$\epsilon=\tau/m=\delta/y$$.
After expansion the above expression retaining again the first three terms
yields:
\begin{eqnarray}
 R_6(n)\simeq\frac{a}{12\tau^3}\frac{m_0^2}{16\tau^2} \Biggl [
-1+\frac{2\delta}{y}\Biggl(4-(n+1)[1+(y+2)/\gamma] \Biggr)
\nonumber\\
-\frac{2\delta^2(n+1)}{y\gamma}
\Biggl((n+2)[2+(y+4)/\gamma+2/(y\gamma)]
-3y-24-(8/3)(2n+5)/y\Biggr)\Biggr]
\end{eqnarray}.

And, finally, for the denominator in (\ref{12}) we obtain
\be K=  \int\limits^{z_m}_0 \frac{z^2 dz}{1+2\varphi/m}e^{-z}
+\frac{a}{12\tau^3}L +
\frac{a}{12\tau^3}\frac{m_0^2}{16\tau^2}(-1+4\delta/y) \ee

\be K^*=  \int\limits^{z_m}_0 \frac{z^2
dz}{(1+2\varphi/m)^2}(1+\frac{4\varphi}{3m})e^{-z}
+\frac{a}{12\tau^3}L +
\frac{a}{12\tau^3}\frac{m_0^2}{16\tau^2}(-1) \ee

resulting in the heavy quark limit in

\be K \simeq D_2 - 2D_3\delta - 4D_4\delta^2+ \frac{a}{12\tau^3}\Biggl(
L -\frac{m_0^2}{16\tau^2}\Biggr)
+\frac{a}{12\tau^3}\frac{m_0^2}{16\tau^2}\frac{4\delta}{y} \ee

\be K^* \simeq D_2 - \frac{8}{3}D_3\delta -  \frac{20}{3}D_4\delta^2+
\frac{a}{12\tau^3}\Biggl( L -\frac{m_0^2}{16\tau^2}\Biggr) \ee

One can easily see, that in the $m\to \infty$ limit $M_n =1$ for
all $n$. This limiting case corresponds to  $c(x)=\delta(1-x)$,
and is accordance with a simple physical picture of the 
infinitely heavy  quark inside heavy hadron carrying the whole
momentum. Interestingly, the  sum rule (12) predict the deviation
from unity, $\Delta=1-M_n$ related with the finiteness of the
heavy quark mass for both  parton distributions and fragmentation
function. From the equations presented before it is easy and
instructive to calculate the actual value of $\Delta$ in the first
and second orders of expansion and compare it with the exact value
of $\Delta$ to examine the accuracy of the expansion in the
inverse heavy quark mass. For numerical analysis we choose
$\bar{\Lambda}=0.6$ Gev, $\omega=1.4$Gev and $\tau=0.6GeV$.  One
can see, that these values reproduce the $D$ meson mass (assuming
$m_c\simeq 1.3 $ GeV), as well as the continuum threshold
$s_0=6\mbox{Gev}^2$, and Borel mass $M_B^2=1-2\mbox{Gev}^2$ in the
ballpark of input parameters used in the sum rules in $D$ channel
(see e.g., \cite{aliev}).  Before we discuss the numerical
results, it should be noted, that the results for moments,
obtained in \cite{n1}, which we have cited before, was obtained in
slightly another way of numerical analysis:- the ratio of the
neighboring moments was considered. As was discuss in \cite {n1},
this two methods give very close results (the difference less than
5-10$\%$, which is within the accuracy of our sum rules approach).

The numerical results for heavy-quark mass expansion of the first
and second moments are shown in Fig.~1 and Fig.~2, respectively,
where they are plotted as a function of the dimensionless ratio
$k=m/m_c$ , for two different values of the ratio
$\gamma=Q^2/M_D^2=2,4$. One can easily see, that even at
relatively large masses of order of the $b$-quark mass ($k\sim 3$)
the heavy mass limit and even the first-order  expansion in
inverse mass are not reliable.  Including the second $O(1/m^2)$
term one improves the situation for the $b$ quark, but the
approximate result for the moments still remains
 very far from  the exact answer in the case of c-quark.
We conclude that the heavy quark limit is not a reliable
approximation  for the parton distributions and fragmentation
functions of c-quark.\\

The author thanks  B.L. Ioffe and A.~Khodjamirian for useful
discussions. This work was supported in part  by the Russian
Foundation of Basic Research, project no. 06-02-16905a an the
funds from EC to the project "Study of the Strong Interacting
Matter" under contract N0. R113-CT-2004-506078.

%\newpage

\newpage

\begin{figure}
\epsfxsize=10cm \epsfbox{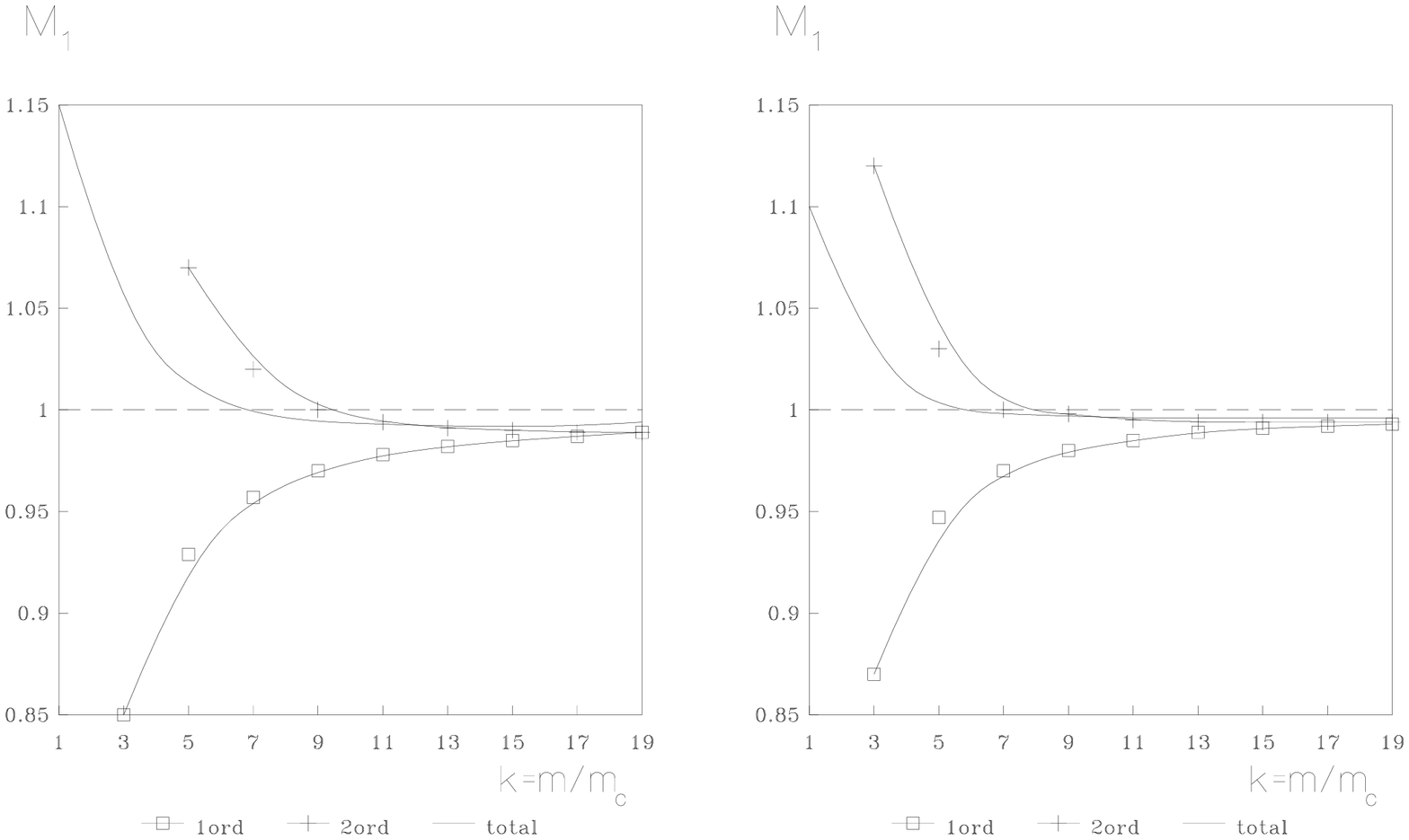}
\caption{ The first moment of the heavy quark distribution function
in pseudoscalar  heavy-light meson plotted as a function of
the quark mass $m$ in the units $k=m/m_c$.
The lines denoted as  "1ord", "2ord" and "total" correspond to
to the first-, second-order expansion in the inverse quark mass
and to the exact answer, respectively. The value of $\gamma=Q^2/M_D^2=2(4)$
is chosen in the left (right) panel.}
\end{figure}

\begin{figure}
\epsfxsize=10cm \epsfbox{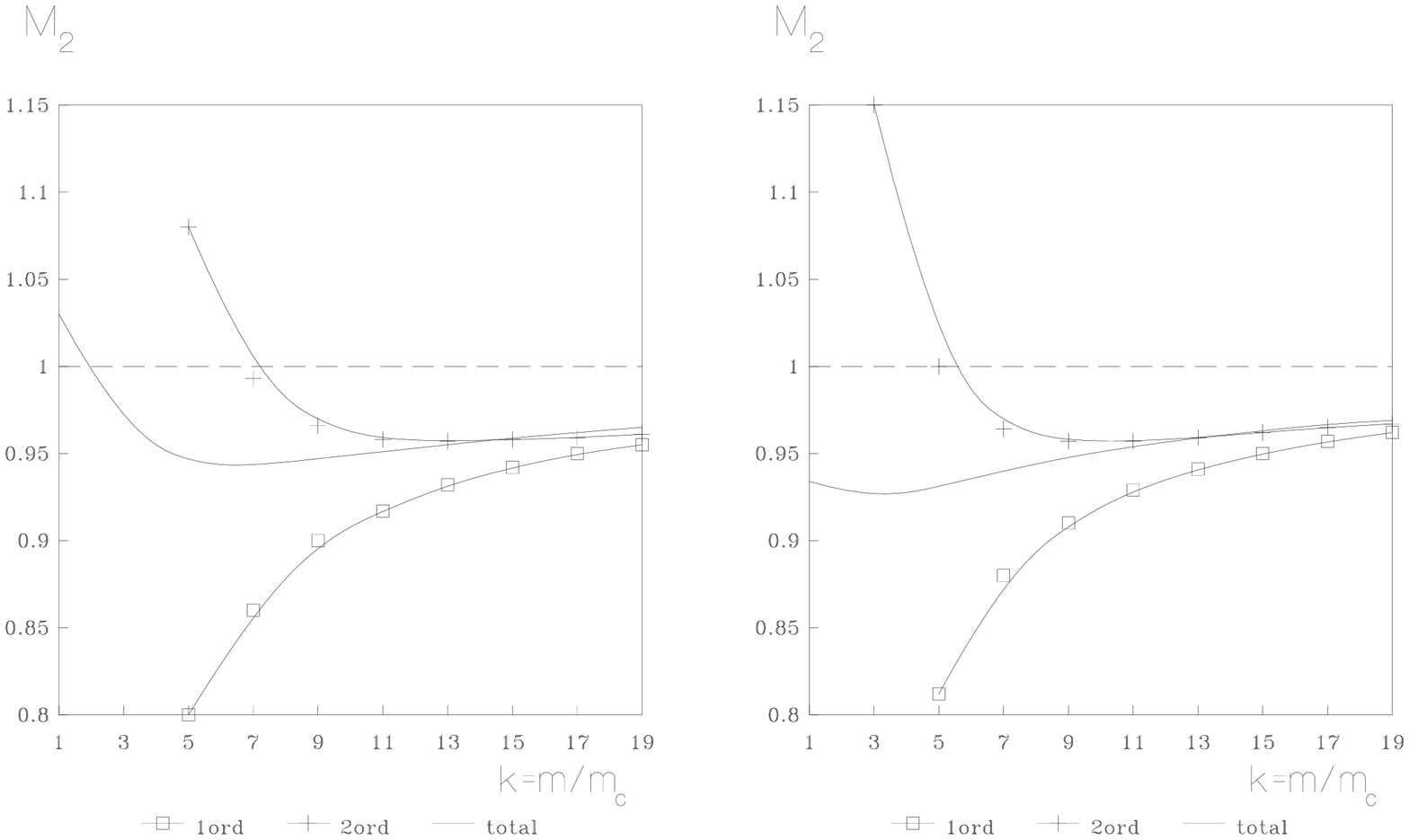} \caption{The same as in Fig. 1
for the second  moment}
\end{figure}

\end{document}